\documentstyle[general,psfig]{mn}  

\title[Galaxy colours in X-ray clusters]{Galaxy colours in high redshift X-ray
  selected clusters -- I: Blue galaxy fractions in eight clusters}

\author[B. W. Fairley et al.]
{B. W. Fairley$^{1}$, L. R. Jones$^{1,5}$, D. A. Wake$^2$, C. A. Collins$^2$, 
  D. J. Burke$^3$, \cr R. C. Nichol$^4$ and A. K. Romer$^4$\\
$^1$ School of Physics and Astronomy, University of
Birmingham, Birmingham, B15 2TT, UK\\
$^2$ Astrophysics Research Institute, Liverpool John Moores University,
Twelve Quays House, Egerton Wharf, Birkenhead, L41 1LD, UK \\
$^3$ Harvard-Smithsonian Center for Astrophysics, 60
Garden Street, Cambridge, MA 02138, USA\\
$^4$ Carnegie Mellon University, 5000 Forbes Ave., Pittsburgh, PA 15213,
USA\\
$^5$ Email: lrj@star.sr.bham.ac.uk\\}
\date{Accepted .....................; Received .....................;
in original form .......................}

\begin{document}                       

\maketitle

\begin{abstract} 

We present initial results from a wide--field, multi--colour imaging project, 
designed to study galaxy evolution in X-ray selected clusters at
intermediate ($z \sim 0.25$) and high redshifts ($z \sim 0.5$).  In this paper we 
give blue galaxy fractions from
eight X-ray selected
clusters, drawn from a combined sample of three X-ray surveys.
We find that all the clusters exhibit excess blue galaxy populations
over the numbers observed in local systems, though a large scatter is present in the
results. We find no significant correlation of blue fraction with redshift at z$>$0.2
although the large scatter could mask a positive trend. We also find no
systematic trend of blue fraction with
X-ray luminosity. We show that the blue fraction is a function of (a)
radius within a cluster, (b) absolute magnitude and (c) the passbands used
to measure the colour. We find that our blue fractions ($\rm f_b$) from galaxy
colours close to restframe 
$(U-B)_0$, $\rm f_b \, \sim \, 0.4$, are systematically
higher than those from restframe $(B-V)_0$ colours, $\rm f_b \, \sim \,
0.2$. We conclude this effect 
is real, may offer a partial explanation of the widely differing levels
of blue fraction found in previous studies and may have implications for
biases in optical samples selected in different bands.

While the increasing blue fraction with radius can be interpreted as
evidence of cluster infall of field galaxies, the exact physical processes
which these galaxies undergo is unclear. We estimate that, in the cores of the more massive
clusters, galaxies should be
experiencing ram--pressure stripping of galactic gas by the
intra--cluster medium. The fact that our low X-ray luminosity systems show
a similar blue fraction as the high luminosity systems, as well as a
significant blue fraction gradient with radius, implies other physical
effects are also important. 
\end{abstract}


\section{Introduction}

During the past few decades, numerous studies of the optical properties
of galaxy populations within clusters have allowed an intriguing insight
into the processes that govern the evolution of galaxies. Rich galaxy
clusters provide a large sample of galaxies, at a similar redshift and of
various masses and morphological type with which to probe these
phenomena. One of the most influential early results, which motivated
many successive studies was that of \scite{but78a}. In this study of two
high redshift clusters they found a large fraction of galaxies in the inner 
regions of the cluster with colours significantly bluer than the
colour--magnitude relation (CMR) of the cluster ellipticals
(Es). \scite{but84} extended this study to include 33 clusters and found
an increasing trend in this blue fraction ($f_B$) with redshift, above $z
\sim 0.1$. A number of subsequent photometric studies have generally
confirmed the presence of these excess blue galaxies, though in varying
quantities (e.g. \pcite{rak95}; \pcite{sma98}; 
\pcite{mar00}; \pcite{kod01}). Subsequently spectroscopic
investigations not only confirmed the cluster membership of large numbers
of these galaxies but also discovered that their spectra contained features 
indicative of star formation, often recently truncated (\pcite{dre85};
\pcite{lav86}; \pcite{cou87}; \pcite{dre92}).

During these studies it became increasingly clear that whilst the luminous
elliptical galaxies present in cluster cores have undergone little but
passive evolution since higher redshift ($z \geq 1$, e.g. \pcite{ell97}; \pcite{sta98}), a major
change has taken place in the other morphological types. Specifically the
large S0 population found in local clusters are found to be almost absent
in high redshift ($z \sim 0.5$) clusters (\pcite{dre97}), where evidence pointed towards
the excess blue galaxies often having spiral morphologies. Confirmation of
this came with morphological studies using $HST$. \scite{dre94} and
\scite{cou94} found that not only were the blue galaxy populations disk
dominated, but they often exhibited signs of interaction. It is now
generally thought that these photometric, spectroscopic and morphological
evolutionary effects are related. The suggestion being that at higher
redshifts, bluer field galaxies, generally spirals, that are drawn into the 
core regions of clusters have their star formation quenched by some mechanism. They
then proceed to evolve morphologies, possibly by some form of mass loss,
into the large S0 population seen in the core of rich clusters
today. Perhaps the most surprising aspect of these phenomena, though, is the
rapid rate at which this evolution takes place.

Clearly understanding the processes behind this evolution is vital to our
understanding of how galaxy histories have progressed. Whilst observational 
data is becoming ever more available, the mechanisms causing this evolution 
are less well understood. Several suggestions have been put forward to
explain the effects of the interaction of galaxies with their environment
and each other. Certainly in the cores of clusters, where the intra--cluster 
medium (ICM) is densest, infalling galaxies, with smaller gravitational
potential wells, will suffer from ram--pressure stripping of their gas
(e.g. \pcite{gun72}; \pcite{aba99}). Other mechanisms which may affect smaller galaxies
include tidal effects generated by the cluster potential (\pcite{byr90}) or 
galaxy harassment caused by close encounters with other galaxies
(\pcite{lav88}; \pcite{moo96a}). In addition to these processes the direct interaction and
merging of galaxies is also a possibility, especially in poorer clusters,
with shallower gravitational potential wells (e.g. \pcite{too72}).

Most studies till now have looked at massive clusters, nearly all of which
have been optically selected. Several large photometric surveys, for
instance, have used optically selected, such as \scite{abe58}, clusters
(e.g. \pcite{but84}; \pcite{rak95}; \pcite{mar00}). Differing blue
fractions, however, were found in these investigations, with \scite{rak95}
claiming a blue fraction of up to 80 percent in high redshift
clusters. Investigations based 
on clusters selected in other wavebands have only recently been
undertaken. \scite{all93} conducted a photometric study of galaxies around
bright radio--galaxies up to $z \, \sim \, 0.45$, and found evidence for enhanced blue
populations in most systems at higher redshift. The galaxy systems observed
contained a number of poorer clusters and no relationship between blue
galaxy fraction and redshift was found. They concluded that the increase in
blue fraction in high
redshift, rich clusters was due to processes inefficient in poorer systems. 

The other area which has only recently begun to be
explored is the selection of clusters, for optical study, via the X-ray
emission from their intra--cluster medium (ICM). \scite{sma98} used a sample 
of 10 intermediate redshift ($z \, \sim \, 0.25$), high X-ray luminosity, X-ray selected clusters
in a multi--colour study of their galaxy populations. They found low
blue fractions, with the bluer
galaxies tending to avoid the cluster 
cores. The other main exploration of optical galaxy properties within
X-ray selected clusters is that of the Canadian Network for Observational
Cosmology (CNOC, \pcite{yee96}). They find excess blue galaxies in
distant clusters (e.g. \pcite{mor98}), colour, emission line, population and age
gradients within the clusters (\pcite{bal97}, \pcite{ell00}), and a fraction of
luminous ``K+A'' cluster galaxies (with strong Balmer absorption lines but no
emission lines) which is similar to that in the field at z$\approx$0.3 (\pcite{bal99}).  
The combined N-body and semi-analytic models of \scite{dia01} match the CNOC1 
cluster observations and support a gradual termination of star formation after cluster
infall. \scite{kod01}, using 7 clusters drawn from 
the CNOC sample also find an excess of blue galaxies. 

Whilst a general consensus
as to the presence of this excess star--formation is evident, the level of
it, and the evolutionary nature is less clear, with a major question being
the steepness of the increase of blue galaxy population with redshift
(e.g. \pcite{mar00}). Blue fraction may, or may not, be a function of
redshift and X-ray luminosity, which may cause significant biases in studies limited to a
small sample of clusters, within a limited range of redshifts or samples
compiled for comparison which have varying X-ray
luminosities (e.g. \pcite{and99}). The works of \scite{lea88}, \scite{sma98} and
\scite{and99} hint at there being no clear relation between blue 
fraction and X-ray luminosity. Additionally, the changing environmental
nature of the clusters may also have a large effect, for instance excess
sub--structure may bias towards higher blue fractions (\pcite{met00}). It
can, therefore, be seen that the large spread in observed results highlights the need for a
method of cluster selection which avoids the biases inherent in optical
selection, whilst still providing large catalogues of galaxy clusters with
multifarious properties to allow exploration of the redshift, cluster
properties and environment parameter space.

This paper, the first in a series, presents details of analysis and initial 
results from multi--colour, wide--field imaging of an X-ray selected sample
of galaxy clusters. The clusters span a large range in X-ray luminosity (a factor of
100 in the 8 clusters presented here) whilst being constrained to two
redshift bands designed to minimise the effects of k--correction. This
project, on completion, should allow an insight into the evolution of
galaxies within a variety of cluster environments, selected solely according to
their X-ray properties. The organisation of the paper is as
follows. In the next section we discuss further the subject of the
selection of the sample of clusters presented here. We then detail the
observations used in this paper, before discussing the reduction and
analysis techniques used. We then present and discuss results from the
photometry, including colour--magnitude diagrams and blue
fractions. Our conclusions are drawn in the final section. In order to
allow comparison to other works we consider an $\rm H_0=50 \: km \: s^{-1} \: Mpc^{-1}$ and $\rm
q_0=0.1$ cosmology throughout this paper.

\section{Sample selection}

\begin{table*}
\begin{minipage}[t]{15cm}
\center{
\begin{tabular}{|c|c|c|c|c|c|} \hline
Cluster & Redshift & $\rm L_x (\rm erg \, s^{-1})$ & Filters & Exposure (ks) & Completeness \\ \hline
RXJ1633.6+5714 & 0.239 & $0.12 \times 10^{44}$ & $B,V,R$ & 3.00, 1.20, 1.20 & 24.3, 24.2, 23.5 \\ 
MS1455.0+2232 & 0.259 & $11.51 \times 10^{44}$ & $B,V,R$ & 2.24, 2.40, 2.70 & 24.0, 23.9, 22.5 \\
RXJ1418.5+2510 & 0.294 & $3.66 \times 10^{44}$ & $B,V,R$ & 3.60, 1.20, 1.20 & 24.2, 23.9, 22.4 \\
RXJ1606.7+2329 & 0.310 & $0.85 \times 10^{44}$ & $B,V,R$ & 3.60, 1.20, 1.20 & 24.2, 24.1, 22.5 \\
MS1621.5+2640 & 0.426 & $4.71 \times 10^{44}$ & $V,R,I$ & 8.40, 5.40, 1.80 & 24.0, 23.9, 22.9 \\
RXJ2106.8-0510 & 0.449 & $2.79 \times 10^{44}$ & $V,R,I$ & 6.90, 4.20, 1.80 & 24.2, 24.1, 22.9 \\
RXJ2146.0+0423 & 0.532 & $4.27 \times 10^{44}$ & $V,R,I$ & 6.00, 2.40, 1.80 & 24.4, 24.2, 22.8 \\
MS2053.7-0449 & 0.583 & $5.49 \times 10^{44}$ & $V,R,I$ & 6.00, 3.90, 1.80 & 24.0, 23.8, 22.8 \\ \hline
\end{tabular}
\caption{\label{tab_obs}Details of the 8 X-ray selected clusters. Redshifts are
    spectroscopically confirmed. X-ray luminosities from the $ROSAT$ PSPC
    are given in the 0.5--2 keV band. Listed are the filters used, along 
    with the respective exposure times. Finally apparent magnitude
    completeness limits are given, estimated from N(m) plots.}} 
\end{minipage}
\end{table*}

As discussed in the previous section, the need for an unbiased cluster
sample with which to investigate galaxy evolution is of paramount
importance. This is certainly the case when trying to construct a picture
of the various mechanisms that affect the changing morphologies and star
formation histories of cluster galaxies. Whilst the studies of optically
rich clusters have given evidence for rapid evolution in galaxy colours,
this is by no means a universal result. Indeed even within the
\scite{but84} analysis, the scatter on the blue fraction versus redshift
plot is large. For instance CL0016+16, A2397 and A2645 (at redshifts 0.541, 
0.224 and 0.246 respectively), to name but a few,
all demonstrate little, if any, evidence
for enhanced blue fraction. This may be indicative of the
fact that optically selected high redshift clusters, which traditionally were often
selected in a blue rest frame (e.g. \pcite{abe58}; \pcite{gun86}), may be an
unrepresentative sample and biased towards high blue fractions. Indeed
\scite{sma98}, using an X-ray selected sample of X-ray luminous massive
clusters found evidence of a spread in blue fractions, though with a low
median value ($\rm f_b=0.04$ for concentrated clusters, for which
\scite{but84} found $\rm f_b=0.09$ for a similar sample), and hence generally low star
formation rates. The goal of this project, therefore, is to construct a
sample of galaxy clusters where the selection of the clusters is
independent of the galaxy properties, with which to investigate cluster
galaxy evolution. Additionally we wish to explore a wide range of cluster masses
(as indicated by their X-ray luminosities) and be able to make comparisons
between similar mass clusters at different redshifts.

The clusters used in this study are drawn from the catalogues of three major 
cluster surveys. The $Einstein$ Medium Sensitivity Survey (EMSS,
\pcite{gio94}) has a higher flux limit than the two other surveys used and
thus primarily contributes brighter X-ray clusters. In addition we use two
other serendipitous cluster catalogues based on archival, $ROSAT$ PSPC
data. These are the Southern--Serendipitous High--redshift Archival $ROSAT$
Cluster survey (S--SHARC, \pcite{bur97a}; \pcite{col97}) and 
the Wide Angle $ROSAT$ Pointed Survey (WARPS, \pcite{sch97};
\pcite{jon98a}). All of these surveys use the X-ray emission from
diffuse gas, trapped in the gravitational potential well of a cluster, as
tracers for the mass of the cluster. Optical imaging and spectroscopic follow--up then
confirms the presence and redshift of the clusters from a few of their brightest
galaxies. This method has the considerable advantage over optical surveys
of nearly eliminating biasing
caused by projection effects and erroneous identifications (\pcite{luc83};
\pcite{fre90}; \pcite{str91}). Whilst the three X-ray surveys used as a
basis for this study vary in their flux limits and detection techniques,
they are all based on the detection of an X-ray flux in a similar energy range,
and they all corrected from measured luminosities to total luminosities in
a similar manner, assuming a King surface brightness profile. 
Mean differences in total flux estimates between X-ray surveys are 
typically $\approx$20 percent (e.g. \pcite{jon98a}, \pcite{vik98b}).

We attempted to restrain the redshift ranges of our cluster sample to
minimise the effects of k-correction between bands. To this end we selected 
two redshift bands for our clusters, in which redshift corrected standard
Johnson--Kron--Cousins filters ($BVR$ in the low redshift sample, $VRI$ in the
high) were mapped onto rest frame $UBV$ filters, as much as possible. The limited number of X-ray selected
clusters in any given RA range, however, prevents a very tight constraint
in redshift range. For the purposes of this initial study, eight clusters were included in our 
sample for observation. These were chosen to maximise the spread in X-ray
luminosity within the constraints placed by the RA range of
this first observing run. Details of these clusters can be found in
Table~\ref{tab_obs}. Here we give cluster IDs along with spectroscopically confirmed cluster
redshifts. All the clusters in this sample have a reasonably relaxed X-ray
morphology as detected in their $ROSAT$ PSPC data. They should, thus, be
good examples of reasonably relaxed systems. The blue fraction of one of
our sample, MS1621.5,
has been studied previously (\pcite{mor98}; \pcite{kod01})
and thus will allow comparison to this investigation, 
providing a good consistency check. The sample presented here is by no
means complete. Future papers will detail observations expanding the sample 
to fill out the redshift and X-ray luminosity parameter space. 

\section{Observations}

The eight clusters detailed in Table~\ref{tab_obs} were observed on June 17--20th 1999
using the Wide Field Camera (WFC, \pcite{ive96}) on the 2.5 metre Isaac Newton Telescope (INT) on La
Palma. The WFC consists of 4 thinned EEV CCDs each having 0.33 arcsec/pixel
resolution and giving a combined 
spatial coverage of $\rm \sim 0.29 \, deg^2$ (each 22.8 
by 11.4 arcmin) at telescope prime focus. The 4 CCDs exhibit varying
characteristics with, in particular, the gain of the central CCD being
lower than the others. In addition there is some non-linearity apparent in all
CCDs, with the most non-linear varying by around 6 percent over the entire
dynamic range. The large sky coverage of the WFC allows us, generally, to image the entirety
of the target cluster in the central CCD. In the cosmology used in this
paper, $\rm H_0=50 \: km \: s^{-1} \: Mpc^{-1}$ and $\rm
q_0=0.1$, the 5.7 arcmin half width of our central CCD corresponds to a
little over 1.7 Mpc, for our nearest cluster, and around 2.9 Mpc for
our most distant. In addition the WFC also provides a highly accurate
statistical estimation of the local background of the cluster, without the
need for separate calibration fields. This will hopefully reduce the affect
of large--scale structural variations on our investigation. We observed
each of the eight clusters in three filters, $BVR$ and $VRI$ for the
low and high redshift clusters respectively. Details of the
exposure times obtained are given in Table~\ref{tab_obs}. In each case we
aimed to achieve good signal--to--noise down to at least $\rm M^* \, + \,
3$ ($\rm M_V \approx -19$). In addition we
give the cluster X-ray luminosity from the $ROSAT$ PSPC (0.5--2
keV). Most observations were carried out near zenith and seeing
conditions were found to be generally good, ranging from 1.52
to 0.79 arcsec FWHM.

\section{Data reduction and Analysis}

\subsection{Reduction and calibration}

\begin{figure}
\psfig{file=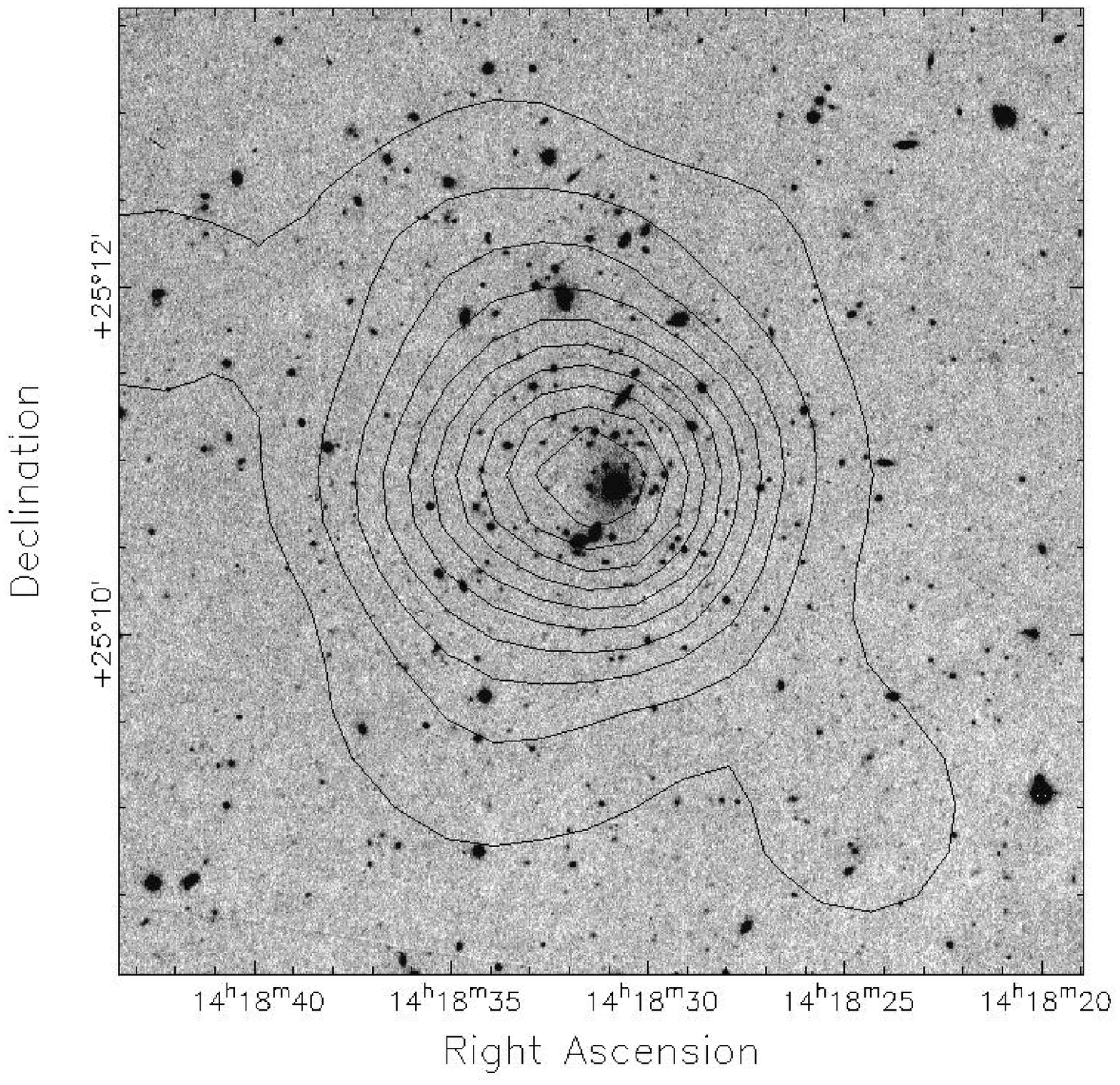,width=10cm,height=7.7cm,angle=270}
\caption{\label{fig_wj1418} INT WFC $V$ band image of RXJ1418.5+2510 at
  $z=0.294$. Overlayed on the image are contours of the X-ray emission
  detected from this $\rm L_x \, = \, 3.66 \, \times \, 10^{44} \, erg \, s^{-1}$
  (0.5-2 keV) cluster from the $ROSAT$ PSPC. Contours are equally spaced in 
  flux level and were derived from an image smoothed to PSPC
  resolution. Only a small fraction ($\sim 2$ percent) of the total WFC
  image is shown.} 
\end{figure}

Initial reduction was carried out in the standard way. After bias
subtraction and non-linearity corrections (obtained from the Wide Field
Survey (WFS, \pcite{mcm99})) were applied, the images were
flat--fielded. Large scale gradients were observed in
the $B$, $V$ and $R$ filters after application of twilight flats, so instead
it was decided to use data flats median combined from the dithered science
images, with masks designed to eliminate bright galaxies and
stars. Due, however, to the low number of target clusters observed during
each night, allied to lengthy exposures (chosen because of long readout times), the
number of images available for median combination each night was
insufficient to generate statistically acceptable flat--fields. It was therefore necessary to combine data from more than one
night. We used median combined data flats combined from two nights, in the $V$ and $R$
bands. In the case of the $B$ band exposures, data was taken on only three
nights, and all were required to provide an adequate flat. In the cases of
the $I$ and $R$ band exposures, night sky fringing was apparent. In the $R$ band
this was found to be virtually eliminated by flat--fielding using data flats. 
The level of fringing in the $I$ band, however, was significantly higher (of
the order of three percent deviation from sky level). The reduction of the
I band thus followed a slightly different pattern. Here we used the
twilight flat--field which produced highly fringed, though generally flat
images. We proceeded to produce a master fringe frame from all of our $I$ band data
and this was used to de--fringe our data. We used an iterative process to
scale the master fringe frame to the approximate level of the fringes;
subtracting this left a residual fringe level in each image, which could be 
used to adjust the subsequent level of the scaling. After final de--fringing, our images
typically showed residual fringing at $\stackrel{<}{_{\sim}} \, 0.3$ percent of sky
level. The above steps were carried out independently on each
CCD. Figure~\ref{fig_wj1418} contains an example $V$ band image of the central region of
RXJ1418.5. Overlayed upon this image are X-ray contours from the $ROSAT$ PSPC.

At least twelve sets of \scite{lan92} standard star fields (Landolt
107,110,113) were observed in various filters, each night, to allow accurate photometric zero--points to 
be obtained for each CCD. Our first night was found to be non--photometric, though in all 
cases clusters observed during this night had short exposure calibration images taken on
one of the following nights. Nights 3 and 4 of the observations appeared to be totally
photometric and the atmospheric absorption on these nights was found to be
consistent with the standard values of the La Palma
site, and thus we adopted these. It was found that an excess amount of atmospheric absorption was
present on the second night ($R \approx 0.30$ mag at zenith). This appeared to be present and
contributing similar absorption in all bands and did not appear to vary
throughout the night. We attribute this roughly grey
absorption to atmospheric dust. It was also observed (with a slightly
larger magnitude ($r' \approx 0.38$)) in the $r'$ band standard star
measurements at the Carlsberg Meridian telescope, on La Palma,  during the
same period. Zero points and colour equations were devised for each CCD, in 
each filter on each night, in order to place the photometry on the standard 
Johnson--Kron--Cousins system. Although the final solutions to these
equations had rms deviations of less than 0.005 mag, the maximum final absolute
photometric error, including systematic effects, was estimated to be
$\stackrel{<}{_{\sim}} \, 0.03$ mag. This estimation was based on standard star deviations from our 
best fit solutions, as well as from external checks which are discussed in
the next section.
In general the solutions were consistent with those
obtained by the Wide Field Survey. In addition zero--points
were adjusted for the affects of galactic absorption, for each cluster
pointing, using absorption maps from the $IRAS$ satellite (\pcite{sch98}).

\subsection{Constructing photometric catalogues}

\begin{figure*}
\begin{minipage}[t]{17.5cm}
\psfig{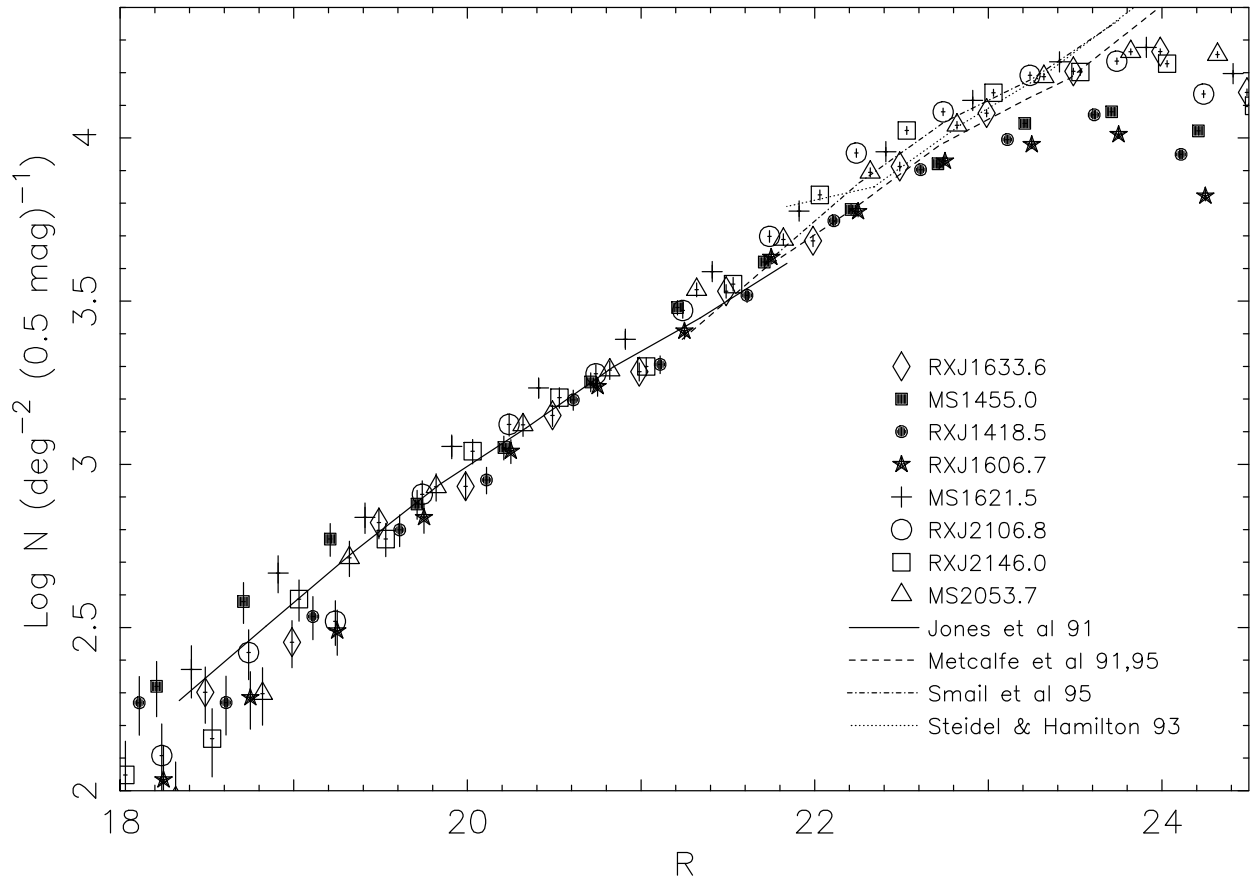}
\caption{\label{fig_back} A composite number--magnitude diagram for the
  combined backgrounds of the clusters in the $R$ band. Different symbols
  represent the 8 different clusters. Error bars are included but are generally 
  smaller than the point size. The lines overlayed are from the
  works of Jones et al. (1991); Metcalfe et al. (1991, 1995); Steidel \&
  Hamilton (1993) and Smail et al. (1995). The turnover due to
  incompleteness generally occurs at brighter magnitudes for the lower
  redshift clusters.} 
\end{minipage}
\end{figure*}

Galaxy photometry catalogues were constructed using the SExtractor 
package (\pcite{ber96}). This allows detection of objects in one image,
with additional photometric analysis from another. Our images were aligned and
trimmed to provide the same sky coverage for all filters for each
CCD. Detection was carried out in the reddest filter image for each
cluster (i.e. $I$ band for the higher redshift clusters, $R$ band for the
lower) so as not to exclude the redder cluster galaxies. Detection significances of more 
than $1.5 \, \sigma$ per pixel were required consisting of at least 9 contiguous
pixels. We used the observed colour of each galaxy, with the appropriate
colour equation, to give accurate magnitudes. We used matching aperture magnitudes for colour estimation, with
the aperture size, a 6 pixel radius equivalent to 2.0 arcsec, larger than the worst seeing in any of our
images. Kron--type magnitudes were measured, using the default
SExtractor settings, to provide the best estimate of the total
magnitudes of the galaxies.

After elimination of objects with unreliable photometry due to, for
instance, proximity to CCD edge, star/galaxy
separation was achieved using the SExtractor CLASS\_STAR parameter estimated
from the two reddest bands for each cluster. At
fainter magnitudes all star/galaxy classification begins to fail, thus we
limited our separation to brighter than an apparent magnitude of $I=22$,
$R=22.5$ and $V=23$. A number of fainter stars will thus be retained, 
however at these fainter magnitudes galaxy counts dominate
(e.g. \pcite{met91}) and the stars will remain in both cluster and field
images and so should be naturally accounted for. In the case of SExtractor
detection in our $I$ band images, a few false detections were noticed by
visual inspection, these coincided with residual fringes in the images. The 
photometry extracted from the other two filters, however, showed no sign of
these detections and so provided a useful robustness check. All galaxies not
detected in at least two of the bands were excluded from our
analysis. This cut may also exclude objects of extreme colour towards our magnitude
limit, these however generally fall well below the magnitude cuts used in
the analysis presented here. 

Number--magnitude plots of galaxies were generated allowing us
to make conservative estimates of the apparent magnitude completeness level of each
observation. An example of one of these plots is shown in
Figure~\ref{fig_back}. Here we plot the number-magnitude relations
for the $R$ band observations of the combined background fields of all our
clusters, in 0.5 magnitude bins. Overlayed on this plot are best fit lines from previous large
scale galaxy number counts investigations (\pcite{jon91}; Metcalfe at al. 1991, 1995;
\pcite{ste93}; \pcite{sma95})\nocite{met91}\nocite{met95}. It can be seen
that, whilst all of our
backgrounds are in reasonable agreement with the previous studies, a significant
spread in  normalisation is found (up to a factor of around 1.5
difference). This is important as it emphasizes the
advantage of having large background areas surrounding the cluster, which
can then be used to take account of variations in the field galaxy density
used for the statistical analysis.
Completeness levels estimated from the number--magnitude plots are also
given in Table~\ref{tab_obs}. The variations between exposure
times and completeness levels are due to the varying observational
conditions. In most cases the completeness levels correspond to
reaching our original target
depth of $\rm M_V \, = \, -19$ (one magnitude deeper than that used in
\scite{but84}) in our cosmology. We miss $\rm M_V \, = \, -19$ by about 0.5 
magnitudes for our most distant cluster, MS2053.7-0449. In addition to the
calibration provided by our standard star measurements we use two other
independent diagnostics to confirm our photometry. The field of
RXJ1418.5+2510 has Johnson $B$--band magnitudes of several stellar sources
published by \scite{tys98}, which agreed to $B$ $>$ 21.5 with our
measured magnitudes, to within 0.03 mag rms. The other checks are of a more
statistical nature. We generated galaxy and stellar count and colour histograms for
several of the background fields of our clusters. These were then compared
to those published by \scite{met91} and again gave good agreement, given
statistical fluctuations. For instance our average ${(B-R)}_{CCD}$ colours of field
galaxies (when converted to the filter colours of \scite{met91}) for our
four low redshift cluster fields were 1.79, 1.75 and 1.79 (in the magnitude
ranges $19.5 \! < \! R_{CCD} \! < \! 20$, $20 \! < \! R_{CCD} \! < \! 
20.5$ and $20.5 \! < \! R_{CCD} \! < \! 21$ respectively) which compare
well with the corresponding values of 1.70, 1.93 and 1.78 (from
\scite{met91}), given the sizeable statistical errors and
large scale structure variations.

\subsection{Galaxy density profiles}

The variation in cluster mass built into our sample means that each cluster 
has a different physical size. This is important to take into consideration 
when comparing the populations of the various clusters. Matched physical
sizes will not provide an accurate comparison, as clusters cores are known
to contain galaxy populations with different characteristics to those in
more outlying regions (e.g. \pcite{dre97}; \pcite{mor98}). A better analysis technique
would be to use the galaxy population as an indicator of the cluster size.

\scite{but78b} detail a method of estimating matched radial distances in
clusters from their galaxy density profiles. They define a concentration
parameter 

\begin{equation}
\rm C = log(R_{60}/R_{20})
\end{equation}

where $\rm R_n$ represents the radius containing n percent of the total
cluster population. For each of our clusters, radial
galaxy density profiles were generated, allowing estimates of the
cluster concentration parameter and the radius of the cluster which
contained 30 percent of the cluster galaxies
($\rm R_{30}$, as used for blue fraction calculations in
\pcite{but84}). 

It was clear from the  statistically background subtracted radial
profiles  where the cluster over-density reached the
background levels, and this allowed estimates of the characteristic
radii from the integrated population within this boundary. 
The large background fields available for each cluster served to
reduce the error in the background level estimates.
The derived values of $\rm R_{30}$ were not sensitive to the details
of this procedure, and for all our clusters $\rm R_{30}$ lay well within the
cluster CCD. Our $\rm R_{30}$ and concentration parameter values for each cluster are detailed
in Table~\ref{tab_clus}. 
It is reassuring to note that all our
clusters have concentration parameters above 0.4, which would be expected
from the relaxed nature of their X-ray contours. It is these concentrated
clusters which were used by \scite{but84} to fit a trend line which defined
the Butcher-Oemler effect.

\subsection{Bright galaxy area correction} 
\label{sec_bcor}

The main
complication of generating accurate radial profiles is the estimation of
the number of fainter galaxies lost behind brighter and larger galaxies in the
foreground. To solve this problem we calculated the total area lost behind
brighter galaxies, in each radial bin used for this calculation. We chose
all galaxies above 300 pixels in size and estimated the area they occupied,
within the radial bin area, at a brightness level above their detection
threshold where blended galaxies would not
have been detected. The area used in the calculations was then corrected
for this lost area. On average the correction was of order one percent for
the field frames and around two percent for the cluster fields,
which agrees well with previous analyses (e.g. \pcite{tre97}).

\section{Results}

\subsection{Colour--magnitude diagrams}

\begin{figure*}
\begin{minipage}{17.5cm}
\psfig{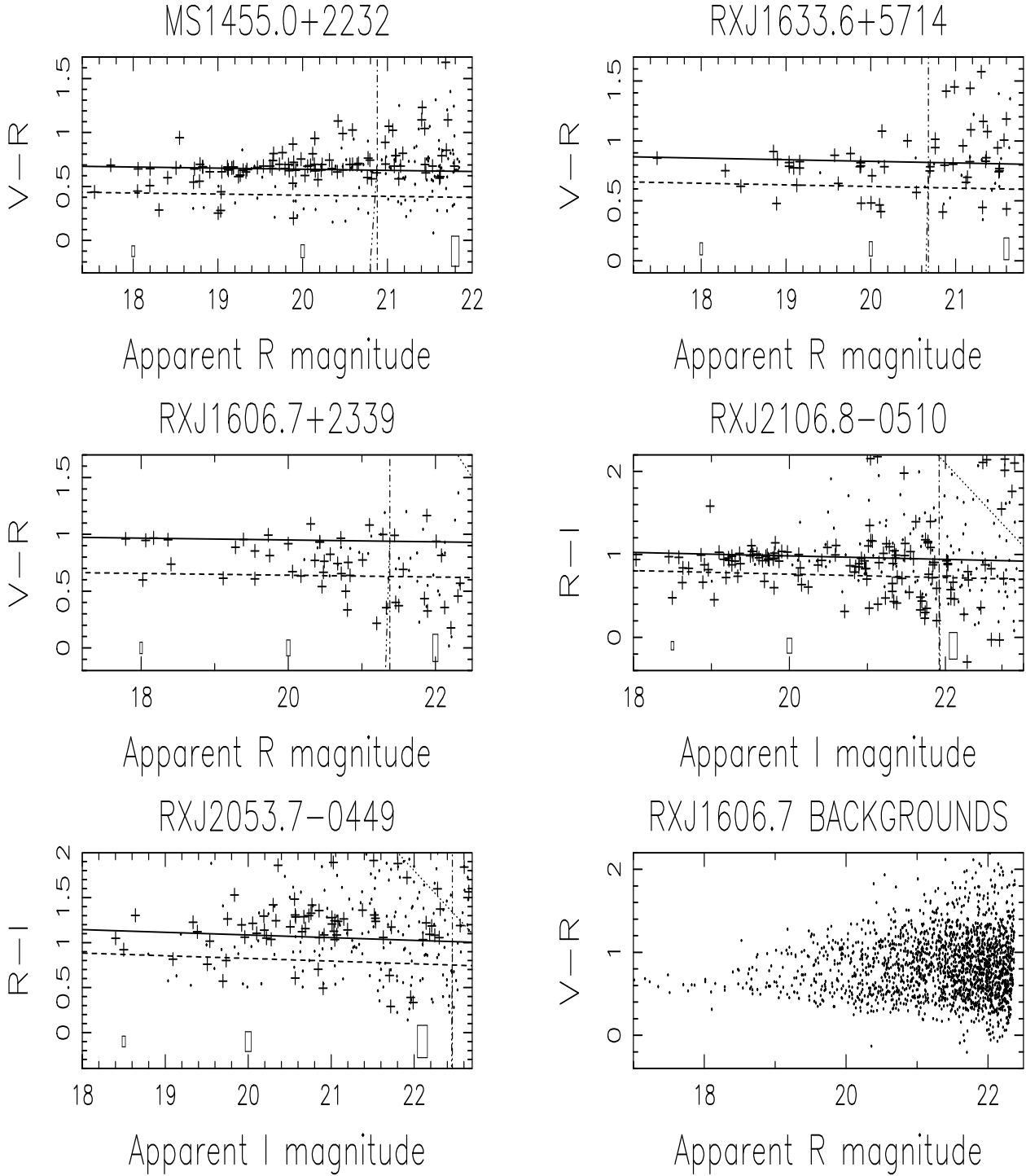}
\caption{\label{fig_cm1} Colour--magnitude diagrams for five X-ray selected
  clusters. Here we plot the colours which correspond approximately to $(B-V)_0$ versus $V_0$
at rest: 
observed $V-R$ against $R$ 
  for MS1455.0+2232 ($z = 0.259$, $\rm L_x = 11.51 \times 10^{44} 
  \, erg \, s^{-1}$), RXJ1633.6+5714 ($z = 0.239$,
  $\rm L_x = 0.12 \times 10^{44} \, erg \, s^{-1}$) and RXJ1606.7+2329 ($z = 0.310$, $\rm L_x = 0.854 \times 10^{44}$),
  and $R-I$ against $I$ for RXJ2106.8-0510
  ($z = 0.449$, $\rm L_x = 2.9 \times 10^{44} \, erg \, s^{-1}$) and
  MS2053.7-0449 ($z = 0.583$, $\rm L_x = 5.5 \times 10^{44} \, erg \,
  s^{-1}$). Also shown is a composite plot of 
  the CM diagrams of the background fields of RXJ1606. 
Only  galaxies within $\rm R_{30}$ are
  shown. Our best fitting CMRs are shown as the solid lines. Blue galaxy cuts used in determining the blue fraction are shown
  as the dashed lines below the CMR line. Vertical dot--dashed lines
  represent a rest frame $\rm M_V \! = \! -20$ magnitude cut, whilst the triple
  dotted--dashed line diverging from this illustrates the effect of the different 
  level of k-corrections for varying galaxy colours. The small boxes at the 
  bottom of each CM diagram represent average colour and magnitude
  photometry error boxes, at the corresponding magnitudes. Finally the plots
   have sloped dotted lines which illustrates if and where our 100
  percent completeness level impinges upon the CM diagram.} 
\end{minipage}
\end{figure*}

\begin{figure*}
\begin{minipage}{17.5cm}
\psfig{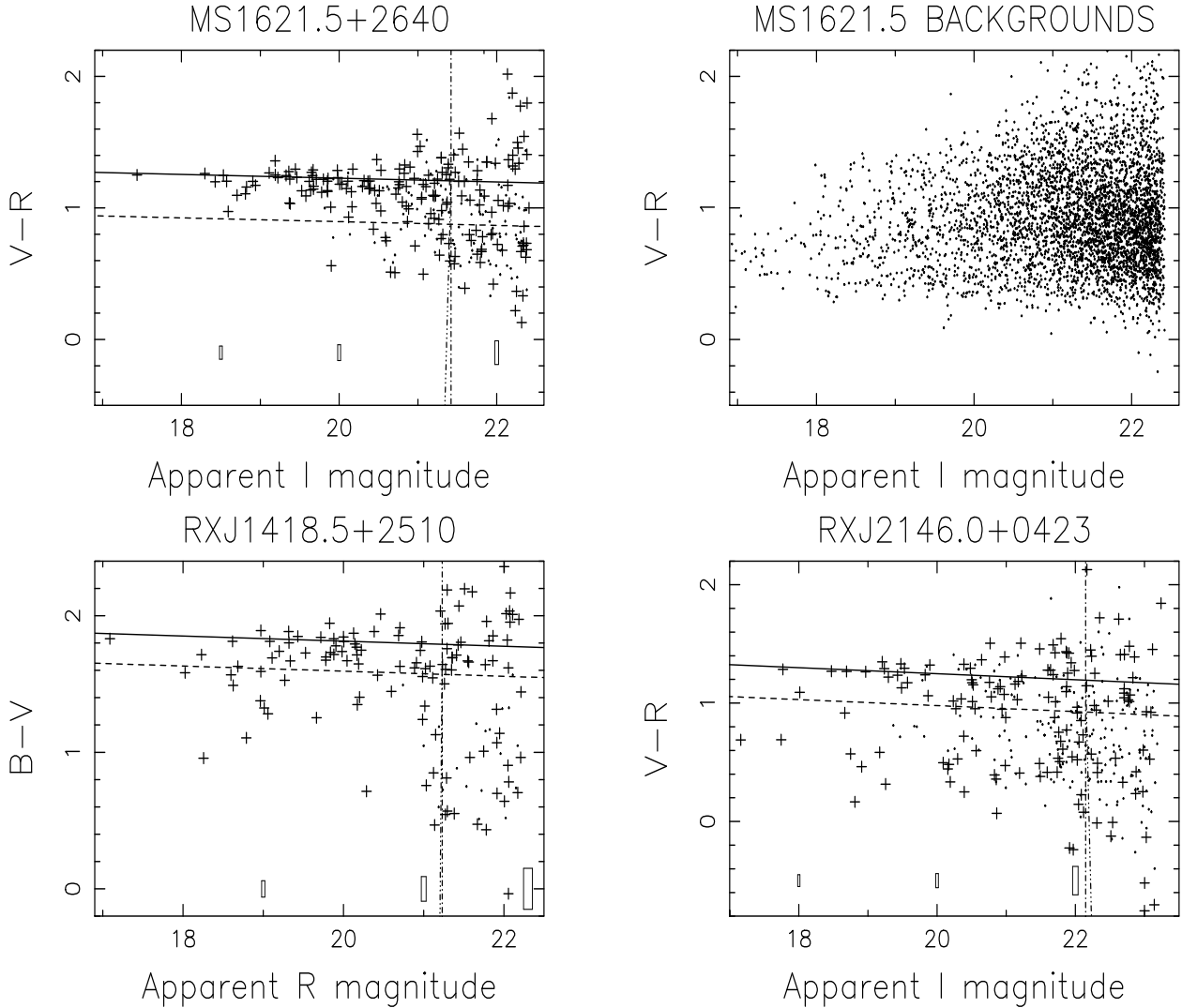}
\caption{\label{fig_cm2} Colour--magnitude diagrams for the three remaining X-ray selected
  clusters. Here we plot the colours which correspond approximately to $(U-B)_0$ versus $V_0$ at rest:
  observed $V-R$ against $I$ 
  for MS1621.5+2640 ($z = 0.426$, $\rm L_x = 4.71 \times 10^{44}
  \, erg \, s^{-1}$) and RXJ2146.0+0423 ($z =
  0.532$, $\rm L_x = 4.27 \times 10^{44} \, erg \, s^{-1}$), and $B-V$
  against $R$ for RXJ1418.5+2510 ($z \, = \,
  0.294$, $\rm L_x = 3.66 \times 10^{44} \, erg \, s^{-1}$).
 Also shown is a composite plot of 
  the CM diagrams of the background fields surrounding  MS1621.5.
Again in all
  these cases galaxies within $\rm R_{30}$ are shown. The lines drawn 
  on the plots represent the same as in the previous figure.}
\end{minipage}
\end{figure*}

Cluster plus field
colour--magnitude (CM) diagrams were generated, along with field CM
diagrams from the three non--cluster CCDs. This allowed a statistical
background subtraction to be carried out, for the purposes of illustration
of cluster against field galaxies. This was achieved by assigning a 
membership probability to each galaxy, determined from the expected number
of galaxies from the background fields. We then used a Monte--Carlo
technique, analogous to that used in \scite{kod01}, to
assign probabilities of cluster membership to each galaxy. The large
background area meant that this could be
achieved in small boxes of size 0.1 mag in colour and 0.5 mag in apparent
magnitude.

In Figure~\ref{fig_cm1} we present five statistically
background subtracted CM diagrams, in which we show colours corresponding
approximately to restframe $B-V$ versus $V$. We plot the apparent $R$ magnitude
against ($V-R$) colour for MS1455.0, an X-ray luminous cluster and
similarly we show ($V-R$) against $R$ for RXJ1633.6 and for
RXJ1606.7, two of our low $\rm L_x$ clusters. We also plot apparent $I$
magnitude against ($R-I$) colour for RXJ2106.8 and MS2053.7, two high redshift clusters. Crosses represent galaxies
statistically flagged as cluster members and dots represent the statistical
field population. Additionally the corresponding
composite field CM plot for the background fields of RXJ1606.7 is
also plotted. 

In Figure~\ref{fig_cm2} we show  CM
plots of the three remaining clusters. 
Here though we plot colours corresponding approximately to restframe
$U-B$ versus $V$. Additionally the corresponding
composite field CM plot for the background fields of MS1621.5 is
shown opposite it. At the bottom of each of these plots are boxes representing 
an average error on our photometry. The errors on the colours were taken
from the error on the aperture photometry and added in quadrature. As can
be easily seen, only in our faintest galaxies does the error on our colours 
begin to reach a level where it would effect our results.

Also shown in Figure~\ref{fig_cm1} are our completeness limits, if they
impinge upon the CM graph. In these cases, only RXJ2106.8 is affected. The
sloped, dotted line in the top--right hand corner represents our
conservative estimate of the 100 percent completeness level. We do not expect this limit to
seriously effect our blue fraction results, at the magnitude
levels of $\rm \sim \, M_V = -20$. Additional evidence for this comes from
the corresponding CM diagram of MS1621.5, from \scite{kod01}, which does not show any
excess red population near our completeness limit.
In the cases of the graphs
displayed in Figure~\ref{fig_cm2}, the completeness limit is harder to
display, as it is a function of all three filter magnitudes. We make an
attempt to constrain these by using colour--colour plots of our galaxies,
which will be presented in more detail in a forthcoming paper. We find that
for our low redshift sample we are 100 percent
complete in $B-V$ at $\rm M_V = -20$, and estimate that this drops
to around 95 percent completeness at $\rm M_V = -19$. For our 
higher redshift sample the results are poorer. Only in MS1621.5 are we
fully complete to
$\rm M_V = -19$ . In the cases of RXJ2106.8 and RXJ2146.0 we are
around 95 percent complete at $\rm M_V = -20$. We are significantly
incomplete to $\rm M_V = -20$ for our furthest cluster,
MS2053.7, and will thus not give results for the $V-R$ blue fraction of this cluster.
The effect of the incompleteness, in the cases where it may impinge on our
CM diagrams, would be to increase the value of the blue fractions
marginally. Our estimates of missing galaxies are confirmed by
comparison with the observations of \scite{met91} who present histogram of
galaxy numbers versus colour. Our field samples are in excellent
agreement with these histograms.

\subsection{Colour--magnitude relation}

The CMR of cluster E/S0s was fitted using a bi--weight estimator
(\pcite{bee90}). In our two poorest clusters, (RXJ1633.6 and RXJ1606.7) we had 
too few cluster galaxies to fit the CMR. Here we took advantage of the
similarity in redshift and X-ray luminosity of these 2 systems. We
combined the CM diagrams of these two systems, correcting for the
difference in E/S0 colour due to redshift (\pcite{fuk95}) and fit for the
CMR from the combined CM diagram. In most cases these bi--weight fits
provided what appeared to be a good estimate of the slope and normalisation 
of the CMR. In certain cases, generally those with poor statistics caused
by lack of galaxies, the CMR fit was not considered to be an accurate
reflection of the actual CMR. In these cases we took a 0.3 magnitude cut
either side of the expected E/S0 colour at $R$ or $I=20$, and
refit the CMR with a weighted least squares fitting algorithm. In general
this served to alter the normalisation of the CMR without altering the
slope significantly. The CMRs of the eight clusters all have a similar slope and are
consistent with a universal CMR, given the errors on the fits. Detailed
analysis of the cluster CMRs will be presented in a future paper.

\begin{table*}
\begin{minipage}[t]{15cm}
\center{
\begin{tabular}{|c|c|c|c|c|c|c|c|c|} \hline
Cluster & Redshift & $\rm L_x (0.5-2 keV)$ & C & $\rm R_{30}$ & $\rm f_b$ &
$\rm f_b$ & $\rm f_b \, (sub)$ & $\rm f_b \, (sub)$ \\
 & & ($\rm erg \, s^{-1}$) & & (arcmin) & ($B-V$) & ($V-R$) & ($B-V$) & ($V-R$) \\ \hline
RXJ1633.6+5714 & 0.239 & $0.12 \times 10^{44}$ & 0.47 & 2.86 & $\rm 0.46 \pm 0.17$
& $\rm 0.25 \pm 0.12$ & $\rm 0.41 \pm 0.14$ & $\rm 0.35 \pm 0.13$ \\ 
MS1455.0+2232 & 0.259 & $11.51 \times 10^{44}$ & 0.62 & 3.75 & $\rm 0.44 \pm 0.09$
& $\rm 0.16 \pm 0.05$ & $\rm 0.34 \pm 0.08$ & $\rm 0.09 \pm 0.04$ \\
RXJ1418.5+2510 & 0.294 & $3.66 \times 10^{44}$ & 0.72 & 2.14 & $\rm 0.37 \pm 0.09$
& $\rm 0.11 \pm 0.05$ & $\rm 0.36 \pm 0.08$ & $\rm 0.12 \pm 0.05$ \\
RXJ1606.7+2329 & 0.310 & $0.85 \times 10^{44}$ & 0.54 & 2.22 & $\rm 0.44 \pm 0.14$
& $\rm 0.20 \pm 0.09$ & $\rm 0.38 \pm 0.11$ & $\rm 0.24 \pm 0.09$ \\ \hline
 & & & & & ($V-R$) & ($R-I$) & ($V-R$) & ($R-I$) \\ \hline
MS1621.5+2640 & 0.426 & $4.71 \times 10^{44}$ & 0.71 & 2.24 & $\rm 0.13 \pm 0.04$ &
$\rm 0.22 \pm 0.05$ & $\rm 0.13 \pm 0.04$ & $\rm 0.23 \pm 0.05$ \\
RXJ2106.8-0510 & 0.449 & $2.79 \times 10^{44}$ & 0.47 & 1.95 & $\rm 0.33 \pm 0.06$
& $\rm 0.26 \pm 0.05$ & $\rm 0.27 \pm 0.05$ & $\rm 0.27 \pm 0.05$ \\
RXJ2146.0+0423 & 0.532 & $4.27 \times 10^{44}$ & 0.49 & 2.30 & $\rm 0.55 \pm 0.08$
& $\rm 0.43 \pm 0.07$ & $\rm 0.44 \pm 0.08$ & $\rm 0.42 \pm 0.07$ \\
MS2053.7-0449 & 0.583 & $5.49 \times 10^{44}$ & 0.46 & 2.08 & Incomplete &
$\rm 0.25 \pm 0.04$ & Incomplete & $\rm 0.20 \pm 0.04$ \\ \hline
\end{tabular}
\caption{\label{tab_clus}Details of the 8 X-ray selected clusters. X-ray
  luminosities from the $ROSAT$ PSPC are given in the 0.5-2 keV
  band. Concentration parameters (C) and $\rm
  R_{30}$ values are derived from background subtracted galaxy radial
  profiles. Blue fractions are given, for $\rm M_v<-20$, in $B-V$ and $V-R$ colours for the low
  redshift sample, and $V-R$ and $R-I$ for the higher redshift
  clusters. Additionally, for comparison, in the final two columns, we give blue fractions derived from CM diagrams
  field subtracted in colour and magnitude boxes (see text).}} 
\end{minipage}
\end{table*}

Finally, to
avoid the problems inherent in k-correction of galaxy magnitudes to
rest frame, we use the approach of \scite{kod01} and convert the
\scite{but84} blue galaxy cut to the observed cluster redshift. This was
done by taking the $\Delta(B-V)_0=-0.2$ colour cut from the CMR
to approximately represent an Sb/Sbc galaxy and estimating, via
interpolation between the different redshift and morphological values, the corresponding colour
difference at the cluster redshift using observed colours (\pcite{fuk95}). 
Overlayed on the CM diagrams in Figure~\ref{fig_cm1} and Figure~\ref{fig_cm2} are the fitted CM
relations along with the resulting blue galaxy colour cut. The graphs
display a magnitude cut equivalent to $\rm M_v=-20$, as used in
\scite{but84}. This was calculated assuming average E/SO colours and in a
cosmology with $\rm H_0=50 \: km \: s^{-1} \: Mpc^{-1}$ and $\rm
q_0=0.1$, and is shown as the vertical dashed--dotted line. This, however,
will tend to include or exclude bluer galaxies, as
their k-corrections are colour dependent. We therefore use observed galaxy
colours (\pcite{fuk95}) to modify this magnitude cut to account for
this. These modifications are shown as the triple--dotted--dashed
lines. Where the $\rm M_v=-20$ line and the colour
corrected magnitude cut line overlap almost completely (e.g in the CM plot
of RXJ1418.5 or RXJ2106.8) we find that the observed filter maps extremely well onto the restframe 
filter for all galaxy colours.

\subsection{Blue fractions}
\label{sec_fb}

Whilst the statistical subtraction of background galaxies from the cluster
plus field CM diagram gives a good indication of the expected distribution
of cluster galaxies within these CM plots, the accuracy is limited by the
size of colour and magnitude bins used. This then may bias any calculation
involving the galaxies selected within these boxes. A better way to utilise 
the large background areas surrounding our clusters would be to directly
calculate the expected number of background blue versus red galaxies, and
then adjust the cluster plus field results accordingly. This avoids any
problems of colour--magnitude bins bisecting the CMR or blue cut line. This
then was the method used for our analysis. We used the largest background
area available, at distances always greater than 7.5 arcmin from the cluster
centre, to calculate the most statistically accurate background
estimate possible. Again the area lost to brighter galaxies was taken
into account during the scaling to the cluster area (see Section~\ref{sec_bcor}). Simple subtraction of
statistical field blue number from actual cluster plus field
blue number, and background total from cluster plus field total
allowed an estimate of the blue fraction to be found. We assumed that the
error on each of these terms could be added in quadrature. Unsurprisingly the
dominating error on the final results comes from the relatively small
excess of blue cluster galaxies.

Table~\ref{tab_clus} shows blue fractions ($\rm f_b$),
calculated at $\rm R_{30}$ and with a magnitude cut of $\rm M_v=-20$ for
our eight clusters. It should be noted that the $V-R$ and $R-I$ blue fractions for our
higher redshift clusters, and the $B-V$ and $V-R$ $\rm f_b$ values for our
lower redshift clusters, roughly correspond to $U-B$ and $B-V$ respectively at
rest. In addition to the blue fractions derived from the method detailed
above, we also compare our estimates to those obtained from the method of
\scite{kod01}. Namely using the colour magnitude diagrams with statistical
background subtraction in colour and magnitude bins. As can be seen from
these figures given in Table~\ref{tab_clus}, we find a general agreement in
results between the two methods, within errors. The other comparison
available to us is, in the case of MS1621.5+2640, with published values of
$\rm f_b$. Whilst our results are not directly comparable to those of
\scite{mor98} and \scite{kod01}, due to the different filters used in the
observations, we find a generally good agreement. Firstly our value of $\rm
R_{30}$, 2.24 arcmin, lies close to
that of 2.18 arcmin as given in \scite{kod01}. Secondly our blue fractions in both 
($V-R$) and ($R-I$), $\rm 0.13 \pm 0.04$ and $\rm 0.22 \pm 0.05$ respectively,
agree with the values of $\rm f_b \, = \, 0.16 \pm 0.04$ and $\rm f_b \,
\sim 0.2$ from the studies of \scite{kod01} and \scite{mor98}
respectively. 

\subsection{Individual cluster anomalies}

Here we briefly discuss issues arising from varying aspects of the
differing properties of our clusters, as well as note any anomalies in our
analysis procedure for individual clusters.

{\bf RXJ2146.0+0423:} The CM diagram for this cluster (see
Figure~\ref{fig_cm2}) shows some evidence of contamination from another
galaxy cluster system. This is evident in what may be a second CMR at
around the $V-R$$\sim 0.6$ level. No X-ray contamination of this cluster
is detected, hence the inclusion of it within our original
sample. Additionally no obvious galaxy excess is visible outside the target 
cluster. This is confirmed by the survey of \scite{gun86} who detect
our target cluster, but find no others within the area of our central
CCD. We note,
therefore, that the estimation of the blue fraction from this cluster is
probably incorrectly high. We also believe this highlights the possibilities of
serendipitous group detection in cluster CM studies. In addition we note
that similar effects are observed in other studies. For instance the CM
diagram of MS1512.4+3647 within the study by \scite{kod01} displays a
similar effect, which the authors attribute to foreground group
contamination.

\begin{figure}
\psfig{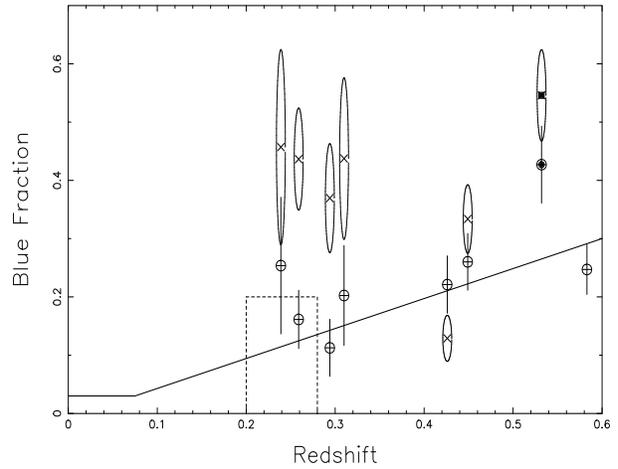}
\caption{\label{fig_bfz} Plot of blue fractions versus redshift for the
  eight clusters presented in this paper. The $\rm f_b$ results presented
  here are calculated using galaxies with magnitudes
  brighter than $\rm M_v=-20$, within $\rm R_{30}$. Open circles (with error bars)
  represent $\rm f_b$ values for red colours ($V-R$ and $R-I$ for low and high
  redshift clusters respectively); diagonal crosses (with error ellipses)
  represent $\rm f_b$ values for blue colours ($B-V$ and $V-R$ for low and high
  redshifts). The solid
  line represents the fit line from Butcher \& Oemler (1984), which should
  be compared with the open circles. RXJ2146.0
  shows signs of foreground contamination (see text) but we include the
  results from this cluster for
  completeness, here represented by the solid circle ($R-I$) and solid square
  ($V-R$). We have also not included the $V-R$ fraction from MS2053.7 as we are 
  incomplete in the $V$ band for this cluster (see text). The dot-outlined
  square represents the blue fraction range from the 10 clusters in the
  X-ray selected sample of Smail et al. (1998). Error bars are one sigma, based on
  Poisson statistics only.} 

\end{figure}

\begin{figure}
\psfig{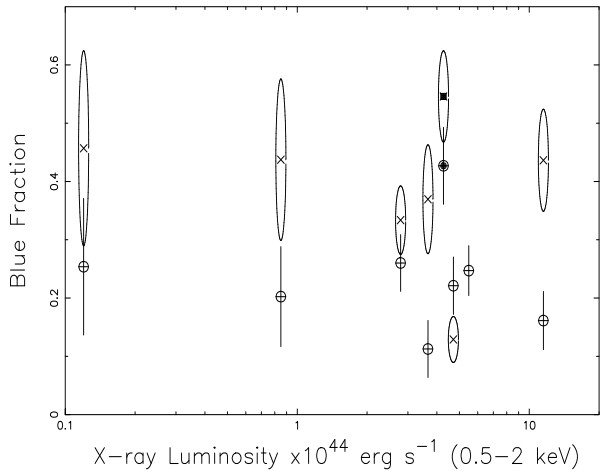}
\caption{\label{fig_bflx} Plot of blue fractions versus X-ray luminosity
  (0.5-2 keV) for the
  eight clusters presented in this paper. The $\rm f_b$ results presented
  here are calculated using galaxies with magnitudes
  brighter than $\rm M_v=-20$, within $\rm R_{30}$. Symbols are the same as in 
  Figure 5:  open circles and vertical crosses (with error bars)
  represent $\rm f_b$ values measured in red colours;
  diagonal crosses (with error ellipses)
  represent $\rm f_b$ values measured in blue colours.
  Again RXJ2146.0 is represented by the solid symbols.} 
\end{figure}

{\bf MS1455.0+2232:} This cluster has the largest calculated value of $\rm
R_{30}$ in our sample (3.75 arcmin) and whilst this radius lies comfortably 
within the central CCD, the radial density profile of this cluster
indicates that the cluster may well spill out into two of the surrounding
CCDs. To negate this problem we use only the furthest CCD from the cluster
(with background area at least 12 arcmin distant) to estimate our field
population. This solution, whilst increasing the statistical error on the
blue fraction estimate, will prevent any cluster contamination of our field 
sample.

{\bf MS2053.7-0449:} As previously noted in Section 4.2 we are not complete
to a level $\rm M_v=-19$ in this cluster. This is primarily due to the fact 
that some of our cluster images were contaminated with multiple tracks
which we believe are reflected light from space debris. Once
eliminated by masks designed for each individual image, this dropped the
detection level below our initial target. We therefore avoid estimating the 
blue fraction at $\rm M_v=-19$, however our estimate of $\rm f_b$ at $\rm
M_v=-20$ in ($R-I$) should still be valid. There may additionally be a low
redshift cluster impinging upon MS2053.7 at about the $\rm R_{30}$
radius. Further calibration images of this
cluster are being obtained and extended results will be presented later in this
series of papers.

{\bf RXJ2106.8-0510:} This cluster is dominated by several large, bright
elliptical galaxies, which in turn serve to obscure a large amount of the
fainter, generally smaller, cluster population. The primary affect of this
is that the estimation of $\rm R_{30}$ is more troublesome than in other
clusters. Additionally the CM diagram is less populated than in other cases, 
resulting in larger errors on the blue fraction.

{\bf RXJ1606.7+2329} and {\bf RXJ1633.6+5714:} Both these clusters are low
X-ray luminosity poorer systems. This tends to increase
the magnitude of the statistical error on the estimate of $\rm f_b$ in both 
cases. Additionally the actual CMR is more difficult to fit (see Section 5.2).

\section{Discussion}

\subsection{Blue fractions against redshift}

We find that most of our clusters have significant blue fractions. In
Figure~\ref{fig_bfz} we plot blue fraction versus cluster
redshift. The open circles represent $V-R$ and $R-I$ $\rm f_b$ values for the
low and high redshift clusters respectively; the diagonal crosses represent 
$B-V$ and $V-R$ $\rm f_b$ values for the low and high redshift clusters.
The solid square
and circle represent the $B-V$ and $V-R$ values of RXJ2146.0+0423 at z=0.532
which suffers from contamination from a foreground group (see previous
section), and will thus have excess blue galaxies. 
We firstly consider the blue fractions measured from the reddest colours,
which correspond approximately to $B-V$ at rest (the circles in
Figure~\ref{fig_bfz}). Although a constant blue fraction of $\rm 
f_b \, \sim 0.2$ is a reasonable description of the current data, the blue fractions are also
consistent with the trend--line of \scite{but84}, which is overlayed on the 
data. A large scatter is also observed, although a large fraction of the
scatter could in principle be due to measurement errors. The dot--outlined
square represents the redshift and blue fraction (estimated using B-I
colours) distribution of the 10 clusters in the \scite{sma98} sample.

It is also
interesting to note that in general there is not a good agreement between
the results from the two colours. In nearly all cases the bluer colour gives
higher $\rm f_b$ values ($f_b \, \sim
0.4$) than the red colour. The reasons for the
excess may be numerous. Firstly the intrinsic range in
galaxy colours is larger in the bluest bands, as illustrated by the
reduction in tightness of the CMR. This also has the effect of making the
blue fraction cut criterion, estimated by the interpolation of predicted
galaxy colours, more difficult to evaluate. Though we estimate
that the colour of the blue galaxy cut would need to be lowered by at least
0.2 magnitudes to decrease $\rm f_b$ from 0.4 to 0.2 as measured in the red 
colour. This is significantly
greater than any error on the interpolation between predicted galaxy
colours. In fainter galaxies 
the scatter in galaxy colours, allied to the larger photometric errors,
may contribute slightly to higher blue fractions. 

We believe, however, that this blue fraction excess can best be explained
in terms of variations in the spectral energy distributions
(SEDs) of the galaxies. The larger range in the blue galaxy colours makes
excess blueness in brighter galaxies (where the
photometric errors are not dominant) much more apparent. The galaxies
contributing to the excess blue fraction
often have $B-V$ colours typical of Sab types whilst having colours
indicative of later type galaxies in $U-B$ (in general
the galaxies that are most blue in one colour are also blue in the other).
This is important when we consider that a fraction of the blue galaxies may
have
starburst or post--starburst SEDs (e.g. have the ``E+A'' or ``k+a'' spectra
of many blue galaxies at high redshift (\pcite{cou87}; \pcite{dre94};
\pcite{cou94})). In fact the observed $B-R$ colours of the blue galaxies in our low
redshift sample agree well with the $B_J-R_F$ colours of the E+A  galaxies
found in \scite{cou87}, when filter conversion is accounted for. 
Spectroscopic observations will be needed to reliably determine the nature of these 
galaxies.

One of the significant implications of the variation in blue 
fraction, as estimated from different colours, concerns sample selection. We
argue that optical selection in any one filter may bias the
cluster sample towards systems more active in the respective filter and
conclude that a method independent of the galaxies, such as X-ray selection,
should be the preferred method of defining a truly representative sample.

\subsection{Blue fractions versus X-ray luminosity}

\begin{figure*}
\begin{minipage}{17.5cm}
\psfig{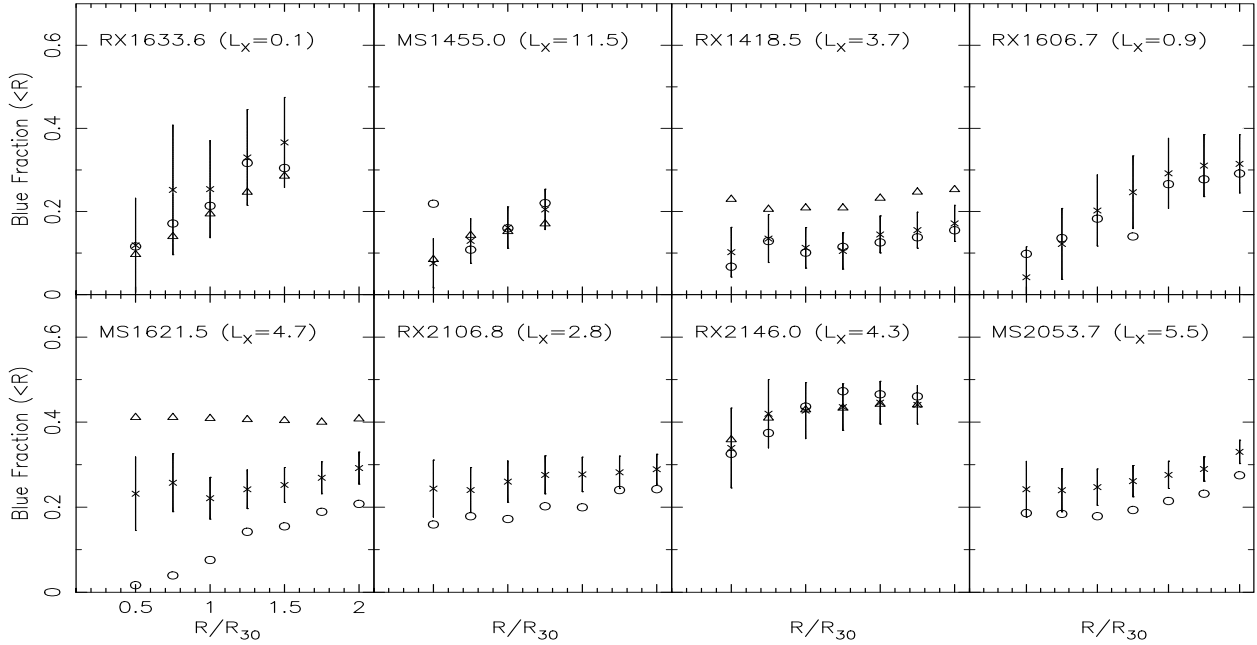}
\caption{\label{fig_fbradbv} Plot of blue fractions versus radius as a
  function of $\rm R_{30}$ for the
  eight clusters presented in this paper. The results shown here are for
  the $V-R$ and $R-I$ colours for the low and high redshift clusters
  respectively, corresponding approximately to $B-V$ at rest. Clusters are
  shown in redshift order, from left to right, with low redshift clusters
  on the top row. The various symbols represent
  differing magnitude cuts. Open circles show blue fractions at $\rm M_v<-21$, diagonal 
  crosses give $\rm f_b$ values for $\rm M_v<-20$ and triangles represent
  $\rm M_v<-19$. We give only results in the magnitude ranges in which we believe we 
  are complete and at radii which are completely enclosed within the
  central CCD. Error bars are given at only one magnitude for clarity. X-ray luminosities are given in units of $\rm 10^{44} \, erg 
  \, s^{-1}$.} 
\end{minipage}
\end{figure*}

\begin{figure*}
\begin{minipage}{17.5cm}
\psfig{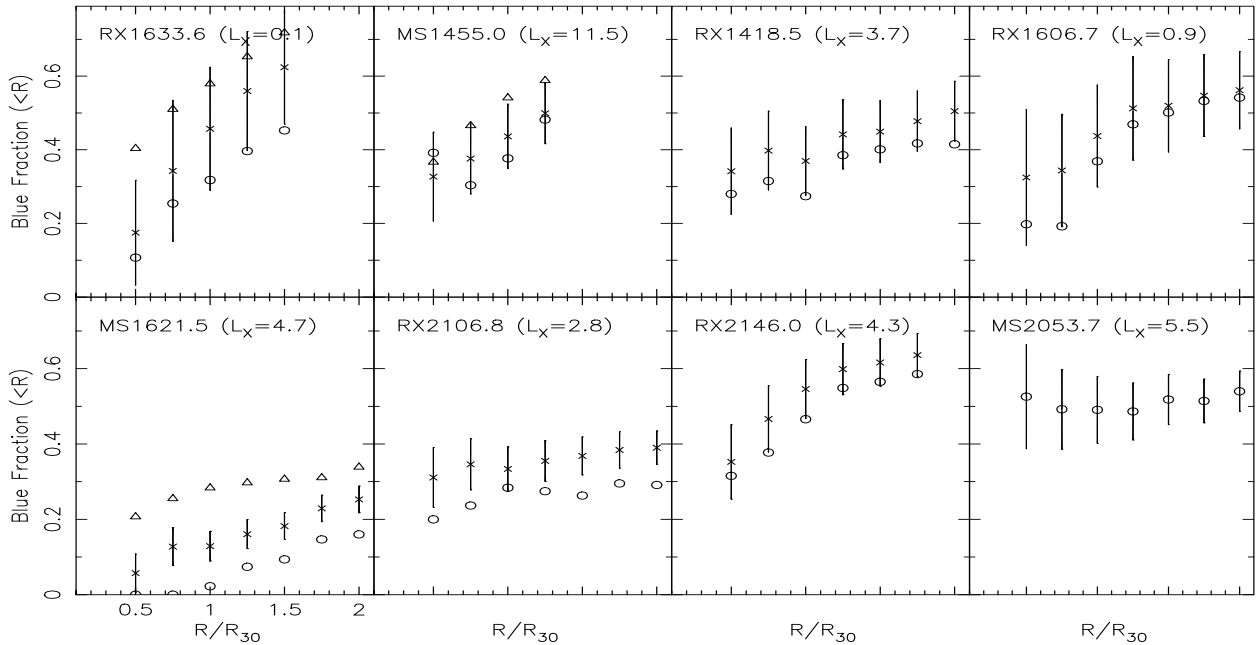}
\caption{\label{fig_fbradub} Plot of blue fractions versus radius as a
  function of $\rm R_{30}$ for the
  eight clusters presented in this paper. The results shown here are for
  the $B-V$ and $V-R$ colours for the low and high redshift clusters
  respectively, corresponding approximately to $U-B$ at rest. Symbols are as
  in Figure 7.} 
\end{minipage}
\end{figure*}

In Figure~\ref{fig_bflx} we investigate
the $\rm f_b$ values as a function of X-ray luminosity. We conclude that no
real trend can be found, over the factor of 100 or so in X-ray luminosity,
though the large errors on the data points makes an expansion of the
sample a priority before any rigorous conclusions can be drawn. This result 
is, however, in agreement with the lack of blue fraction correlation to X-ray 
luminosity in the \scite{but84} sample (\pcite{lea88}; \pcite{and99}). If this
result is substantiated, it has important implications for the physical
processes that are responsible for the Butcher--Oemler effect. 

An interesting 
question is whether the galaxies within $\rm R_{30}$ would expect to
have had their gas removed via ram--pressure stripping due to the ICM
(\pcite{gun72}). 
An estimate of the dependence of ram--pressure stripping with $L_x$
can be made as follows.
The rate of ram--pressure stripping is
proportional to $\sim \, {\rho}_{\scriptscriptstyle{ICM}} \, {V_{gal}}^{2.4}$ 
(\pcite{gae87}) and 
$L_x \, \propto \, {{\rho}_{\scriptscriptstyle{ICM}}}^2$.  
From the Virial theorem and the luminosity--temperature relation, $V_{gal} \,
\propto \, T^{0.5} \, \propto \, {L_x}^{1/6}$, and so the ram--pressure
stripping rate will increase as $\sim \, L_x$. Thus the 
truncation of star--formation by ram--pressure stripping is
expected to be significantly more effective in high $\rm L_x$ clusters,
especially over the factor of 100 in $\rm L_x$ sampled here.

\scite{ste99} model a large spherical galaxy ploughing through a dense
ICM and being stripped of gas. They find that gas stripping is primarily
a function of cluster temperature (and hence mass), galaxy velocity within
the ICM, the gas replenishment rate from stars and the distance from the
centre of the cluster potential. We can make an estimate
of cluster mass from our X-ray luminosities via the X-ray
luminosity--temperature relation (e.g. \pcite{dav93}) which has been shown
to be non-evolving for the redshift ranges of our clusters (\pcite{mus97};
\pcite{fai00}). Our most massive cluster, MS1455.0, in which ram--pressure
stripping should be most powerful, should have a temperature of
approximately 6 keV. We estimate, from predictions presented 
in Figure 6 of \scite{ste99}, that 
ram--pressure stripping will only be a dominant force towards the centre of
clusters, and then only for luminous, high temperature clusters (i.e. $\rm
T_x \, \stackrel{>}{_{\sim}} \, 3 \, keV$; see also \pcite{aba99}). 
Thus, even if the orbits of most galaxies within $\rm R_{30}$ have taken them close to the
cluster centre in the past, we would not expect ram-pressure stripping  to be
a dominant process in the low $\rm L_x$ clusters. To increase the 
ram-pressure force, we would
require that
either infalling galaxies have velocities significantly larger than the
mean galaxy velocity (which is not inconceivable, e.g. \pcite{hen87}), or
that the gas replenishment rate is too low to be effective in replacing
stripped gas, which may be less plausible. 
If the excess of blue galaxies 
in the low X-ray luminosity clusters studied here contains a large fraction 
of post--starburst galaxies, as found in richer clusters, then a mechanism
other than ram--pressure stripping is probably responsible for truncating
the star formation.

\subsection{Blue fractions with radius and galaxy infall}

It has been widely noted that blue fraction varies with cluster--centric
radius. To investigate this in our sample we estimate the blue fractions
detailed in the previous sections within various radii for each cluster. We
consider here both colours and study the variation of $\rm f_b$ with
radius as a fraction of $\rm R_{30}$. Figures~\ref{fig_fbradbv} and
\ref{fig_fbradub} show plots of 
integrated blue fraction versus radius, for each of our
clusters, in the red and blue colours respectively. We plot the blue fraction at varying magnitude cut--offs, in order to
estimate the effect of including brighter or fainter galaxies. As can be
clearly seen there is a large variation in the
profiles of some of the clusters, a small change in others. In most of the
clusters we see an increasing blue fraction with
fainter magnitude cut (in agreement with \scite{kod01}). This trend,
however, is not universal in our cluster sample. We note also that the
slope of $\rm f_b$ with radius, for each cluster, is similar in both colours, 
with the principle difference again being the magnitude of $\rm f_b$.

Most of the clusters
exhibit at least some increase in blue fraction with radius, which highlights the need for caution when
interpreting a blue fraction, at a characteristic radius, for a
cluster. However the main conclusions drawn in the previous sections should 
not vary dramatically given small shifts in characteristic radii. The radius used here, $\rm R_{30}$, is of course
dependent on the galaxy distribution. We also investigated the effect of
using a radius derived from the X-ray properties of the cluster on the blue 
fractions. Here we used our estimated cluster temperatures to assign a
cluster radius ($\rm r_{200}$), assuming the clusters followed an NFW profile
(\pcite{nav95}). We found that the blue fractions at $\rm r_{200}/2$ were
in agreement (within errors) with those at $\rm R_{30}$. This is not too
surprising as the galaxy population would be expected to roughly trace the
cluster dark matter profile in relaxed clusters. 

To what extent are the conclusions drawn from Figures 5 and 6 dependent on the
absolute magnitude limit and the radius used to measure the blue fraction? 
Figures 7 \& 8 show that the broad trends of the variation of blue fraction with 
absolute magnitude, radius and colour are similar from cluster to cluster, so that
the conclusions drawn from Figures 5 and 6 are to a large degree independent of the precise
choice of absolute magnitude limit or radius,
so long as a consistent choice (with radius as a fraction of $\rm R_{30}$) is made. 

The increase in blue fraction with radius in many clusters suggests that it 
is field galaxies falling into the cluster
which cause the enhanced blue population (e.g. \pcite{ell00}) and which
may be one of the prime causes of the well known morphology--density
relation, where spiral fraction increases with decreasing galaxy number
density (e.g. \pcite{dre97}). The mechanism by which the infalling field
galaxies are morphologically transformed and have their star formation
truncated could be ram--pressure stripping. As
discussed in the previous section, the likelihood of ram-pressure stripping
increases with ICM density towards the cluster centre. Gas stripping, however, should be less 
important in the lower luminosity clusters (e.g. RXJ1633.6 and RXJ1606.7). Thus
the increase in blue fraction with radius seen in these systems should not
be due to ram-pressure stripping.

\section{Conclusions}

We have presented the first results from a large multi--colour, wide--field
imaging program of high redshift X-ray selected clusters. We use the large
background area available from our observations of 8 clusters, using the INT
WFC, to allow a statistical estimate of the fraction of blue galaxies in
the cluster cores. The $B$, $V$, and $R$ band
photometry for four intermediate redshift clusters ($z \sim 0.25$) and $V$, $R$ and $I$ band photometry
for four higher redshift systems ($z \sim 0.5$), corresponded approximately
to $UBV$ at rest.

We find no significant change of blue fraction with redshift in the range 
0.2$<$z$<$0.5, but the  blue fraction in all the clusters is higher than in
those of \scite{but84} at z$<$0.1.  Our reddest colour matches that used by \scite{but84}.
In this colour, the blue fractions we measure are equally 
consistent (within the measurement errors) with  the trend--line presented in 
\scite{but84}, or with no trend in  blue fraction with redshift at 0.2$<$z$<$0.5.
Our blue fractions in this colour are also in agreement with those measured in X-ray selected,
luminous clusters by \scite{kod01} over a similar redshift range as used here.
However, our blue fractions are higher than those measured in X-ray selected,
luminous clusters by \scite{sma98} at z$\approx$0.25. \scite{sma98} find a median
blue fraction of $f_b=0.04\pm0.02$ for concentrated clusters ($\rm C>0.35$).
A possible explanation lies in the different colour (restframe $U-R$ or observed $B-I$)
and selection band (restframe $R$) used by \scite{sma98} to define the blue
fraction. 

We note a significant
difference in the blue fraction values as
calculated in our two different colour bands. This may explain some of the
scatter in previous blue fraction measurements, which use a variety of
different colours, and may additonally imply some level of bias in $\rm
f_b$, depending on the optical band used for cluster selection.

We find no evidence for a trend in blue fraction with
X-ray luminosity, though again our results exhibit a large
scatter. Interestingly
we do not find the relation between blue fraction and richness that
\scite{mar01} find. Unfortunately, whilst that study was based on a large
cluster catalogue the authors chose to extract blue fractions at a common fixed
spatial radius for each cluster. This obviously prevents direct comparison
with this work. It may also explain their richness correlation result,
since a lower blue fraction in richer clusters could be due to the smaller
radii (relative to $\rm R_{30}$) used to sample the richer clusters,
combined with the morphology--density relation.

 Profiles of blue fractions with radius show an increasing proportion
of blue galaxies, often faint, towards the outskirts of the clusters. The
fact that the blue fraction is a function of radius, of faint magnitude cut,
and also of the colour used to define it, emphasizes the need
for extreme caution when assigning a characteristic blue fraction for an
individual cluster. This may provide an explanation for the widely
differing blue fraction levels reported in various studies. 
\scite{rak95}, for example, find a high blue fraction of 0.8 at z=0.9, but use a 
different method to most studies, employing Stromgren photometry, a variable
cluster radius and photometric redshifts to help remove background galaxies.
Comparison of blue fractions measured using different techniques may not be valid.

Our results suggest that the increased blue fractions are caused by infalling
field galaxies, which may or may not undergo a starburst, before having
their star formation truncated. We would expect that ram--pressure
stripping occurs in the cores of our more X-ray luminous clusters,
and this would cause the lowering of blue fraction, towards the cluster
cores, that is observed. This explanation, however, is less likely to
explain the same observational result in lower $\rm L_x$ systems, where
other physical processes (such as galaxy interactions) must be occurring. Our 
expanded cluster sample will allow a better understanding of the physical
processes giving rise to blue galaxies in high redshift clusters.

\section{Acknowledgments}

We thank Ale Terlevich for his help with the CMR fitting and are grateful
for useful discussions with Dave Gilbank regarding INT WFC
reduction. De-fringing of data was carried out using the code developed by
Mike Irwin and we
would also like to thank the Cambridge CASU for providing public access to the
WFS calibration details on their website
({http://www.ast.cam.ac.uk/$\sim$wfcsur/}). We thank the referee for useful 
comments. 
Computing facilities provided by the STARLINK
project have been used in this work. BWF and DAW acknowledge the receipt of
PPARC studentships and LRJ also acknowledges the support of PPARC. DJB
acknowledges the support of SAO contract SV4-64008.

\bibliographystyle{mnras}
\bibliography{clusters}

\end{document}